\documentclass[twocolumn,aps,footinbib,superscriptaddress]{revtex4-2}
\usepackage{latexsym}
\usepackage{graphicx}
\usepackage{bm}
\usepackage{braket}
\usepackage{amsmath}
\usepackage{amssymb}
\usepackage{hyperref}
\usepackage{mathrsfs}
\usepackage{cancel}
\usepackage{color}
\usepackage{graphicx}
\usepackage{multirow}
\usepackage{mathtools}
\usepackage{mathrsfs}
\usepackage{comment}
\usepackage{cancel}
\usepackage{pst-3dplot}
\usepackage{pst-solides3d}
\usepackage{pst-all}
\renewcommand{\d}{{\rm d}}
\renewcommand{\i}{{\rm i}}
\newcommand{\e}{{\rm e}}

\newcommand{\sgn}{{\rm sgn}}

\renewcommand{\Im}{{\rm Im}\;}
\usepackage{xcolor}
\hypersetup{
    colorlinks=true,
    citecolor=[rgb]{0,0,0.55},
    linkcolor=[rgb]{0,0,0.55},
    urlcolor=[rgb]{0,0,0.55},
}
\usepackage{xr}
\usepackage{version}	% required for `\comment' (yatex added)
\begin{document}
\title{Chern numbers in two-dimensional systems with spiral boundary
conditions}
 \author{Masaaki Nakamura and Shohei Masuda}
\affiliation{
Department of Physics, Ehime University Bunkyo-cho 2-5, Matsuyama, Ehime
790-8577, Japan}
\date{\today}
\begin{abstract}
We discuss methods for calculating Chern numbers of two-dimensional
lattice systems using spiral boundary conditions, which sweep all
lattice sites in one-dimensional order. Specifically, we establish the
one-dimensional representation of Fukui-Hatsugai-Suzuki's method, based
on lattice gauge theory, and the Coh-Vanderbilt's method, which relates
to electronic polarization. The essential point of this discussion is
that the insertion of flux into the extended one-dimensional chain
generates an effective current in the perpendicular direction. These
methods are valuable not only for a unified understanding of topological
physics in different dimensions but also for numerical calculations,
including the density matrix renormalization group.
\end{abstract}

\maketitle

%%%%%%%%%%%%%%%%%%%%%%%%%%%%%%%%%%%%%%%%%%%%%%%%%%%%%%%%%%%%%%%%%%%%%%%
%% Introduction
%%%%%%%%%%%%%%%%%%%%%%%%%%%%%%%%%%%%%%%%%%%%%%%%%%%%%%%%%%%%%%%%%%%%%%%
\section{Introduction}
\label{sec:intro}

% topological physics
For over a decade, there has been extensive research on topological
phases and transitions in the context of topological insulators
\cite{Haldane1988,Kane-M,Bernevig-Z,Bernevig-H-Z2006,Ryu-S-F-L,
Konig-2007,Konig-2008}. Topological insulators exhibit energy gaps in
the bulk and gapless edge (surface) states in two (three)
dimensions. The concept of topology has been further extended to include
topological superconductors, Kitaev systems, non-Hermitian skin effects,
and other related phenomena \cite{Read-G,Kitaev2006,Yao-W}.

% 1D topology
On the other hand, the study of topological phases and transitions in
one-dimensional (1D) quantum spin systems has been ongoing since the
1970s. For instance, the dimer-N\'eel transition in a spin-$1/2$
frustrated anisotropic Heisenberg chain and the successive dimerization
observed in Haldane gap systems are interpreted as transitions between
different gapped phases that possess distinct topological properties
\cite{Haldane1982,Haldane1983a,Haldane1983b,Affleck-H}.
%

% polarization
In this paper, we present a unified study of topological phases and
transitions in various systems and dimensions using the concept of
electronic polarization and related concepts \cite{Resta1994, Resta,
Resta-S1999, Resta2000}. In 1D lattice electron systems, we introduce
polarization operator (or twist operator) $U$ defined as the exponential
of the position operator, and consider its expectation value in the
ground state $\ket{\Psi_0}$:
\begin{equation}
 z=\braket{\Psi_0|U|\Psi_0},\quad
  U=\exp\biggl(\i\frac{2\pi}{L}\sum_{j=1}^L j n_j\biggr).
  \label{def_z}
\end{equation}
Here, $L$ represents the number of sites, $n_j$ is the electron number
operator at the $j$-th site. Resta established a relationship between
$z$ and electronic polarization, given by
$\lim_{L\to\infty}(e/2\pi)\Im\!\ln z$ \cite{Resta}. The signs of $z$
provide information about the system's topology, such as the presence of
charge or spin density waves and edge states
\cite{Nakamura-V,Nakamura-T}. Furthermore, the condition $z=0$ can be
utilized to detect phase transition points.

% relation to Lieb-Schultz-Mattis theorem
The same quantity as in Eq.~(\ref{def_z}) was also introduced in the
Lieb-Schultz-Mattis (LSM) theorem for 1D quantum systems
\cite{Lieb-S-M,Affleck-L,Affleck,Oshikawa-Y-A,Yamanaka-O-A}.  In the LSM
theorem, Eq.~(\ref{def_z}) appears as an overlap between the ground
state and a variational excited state. According to the theorem, an
energy gap above non-degenerate ground state is possible for $z\neq 0$
with $L\to\infty$.

\begin{figure*}[t]
\begin{center}
 \includegraphics[width=18cm]{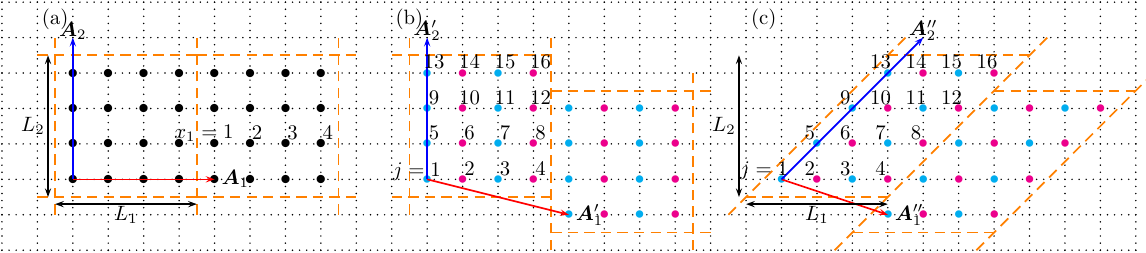}
\end{center}
 \caption{Several boundary conditions for 2D square lattices with
 $L_1\times L_2$ sites: (a) Conventional periodic boundary conditions
 (PBCs), (b) Spiral boundary conditions (SBCs) for $(\pi,0)$ charge
 order, and (c) SBCs for $(\pi,\pi)$ order. For the systems with SBCs,
 the lattices are labeled as extended 1D chains.  }
\label{fig:unit_SBC}
\end{figure*}

% extention to 2D
Thus the property of $z$ has been well studied for 1D systems. However,
its application to higher-dimensional systems is not fully discussed.
In our preceeding work \cite{Nakamura-M-N}, we have extended the twist
operator in Eq.~(\ref{def_z}) to two-dimensional (2D) systems using
spiral boundary conditions (SBCs) that encompass all lattice sites in
one-dimensional order as illustrated in Fig.~\ref{fig:unit_SBC}.  These
boundary conditions have been utilized to extend the LSM theorem to
higher dimensions \cite{Oshikawa-2000a,Oshikawa-2000b,Hastings},
eliminating unphysical limitations on the system size \cite{Yao-O}. By
calculating $z$ with SBCs, we have examined the 2D Wilson-Dirac model
\cite{Wilson,Qi-W-Z} and identified topological transition points as
$z=0$.

% transverse response
However, the aforementioned extension does not account for transverse
response, so that it does not include the information of the Chern
number. Therefore, $z$ needs to be modified to incorporate the flux
effect, which is perpendicular to the twist direction. For conventional
periodic boundary conditions (PBCs), this extension was carried out by
Coh and Vanderbilt for the Haldane model \cite{Coh-V}.  On the other
hand, Fukui, Hatsugai, and Suzuki proposed a method to calculate the
Chern number based on the lattice gauge theory
\cite{Fukui-H-S}. Therefore, we aim to elucidate the relationship
between the Coh-Vanderbilt's method and the Fukui-Hatsugai-Suzuki's
method. Subsequently, we derive the formalism of electronic polarization
with SBCs to identify the system's topology by Chern numbers.

% organization of this paper
This paper is organized as follows. In Sec.~\ref{sec:Chern_number}, we
provide a review of the two methods for calculating Chern numbers in
lattice systems. Next, in Sec.~\ref{sec:SBC}, we discuss the
justification of twist operators with SBCs and establish the
correspondence between wave numbers in conventional PBCs and those in
the SBCs. In Sec.~\ref{sec:Results}, we employ the two methods with SBCs
to calculate Chern numbers in several models. Finally, we present a
summary and discussion in Sec.~\ref{sec:Summary}. In
Appendix~\ref{sec:Berry_con}, we examine the Chern numbers obtained by
the two methods in the continuum limits. Throughout this paper, we set
the lattice constant $a$ and the reduced Planck constant $\hbar$ to
unity.

%%%%%%%%%%%%%%%%%%%%%%%%%%%%%%%%%%%%%%%%%%%%%%%%%%%%%%%%%%%%%%%%%%%%%%%
%% Chern number
%%%%%%%%%%%%%%%%%%%%%%%%%%%%%%%%%%%%%%%%%%%%%%%%%%%%%%%%%%%%%%%%%%%%%%%
\section{Chern number}
\label{sec:Chern_number}

In this section, we will review two methods for calculating the Chern
number in 2D lattice systems: the Fukui-Hatsugai-Suzuki(FHS)'s method
and the Coh-Vanderbilt(CV)'s method. The CV's method is specifically
related to the electronic polarization.

%%%%%%%%%%%%%%%%%%%%%%%%%%%%%%%%%%%%%%%%%%%%%%%%%%%%%%%%%%%%%%%%%%%%%%%
%% Fukui-Hatsugai-Suzuki's method
%%%%%%%%%%%%%%%%%%%%%%%%%%%%%%%%%%%%%%%%%%%%%%%%%%%%%%%%%%%%%%%%%%%%%%%
\subsection{Fukui-Hatsugai-Suzuki's method}
\begin{figure}[b]
\begin{center}
 \includegraphics[width=6cm]{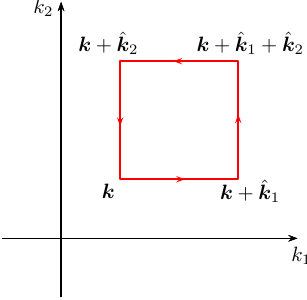}
\end{center}
\caption{Path along a plaquette for the Fukui-Hatsugai-Suzuki's method.}
 \label{fig:plaquette}
\end{figure}

A method for calculating the Chern number in lattice systems has been
proposed by Fukui, Hatsugai, and Suzuki \cite{Fukui-H-S}.  In this
method, we define the following function as an overlap between
neighboring two wave vectors:
\begin{equation}
 V_{\bm{k},\bm{k}+\hat{\bm{k}}_{\mu}}
  =\braket{u(\bm{k})|u(\bm{k}+\hat{\bm{k}}_{\mu})},
\label{U(1)_link}
\end{equation}
where $\hat{\bm{k}}_{\mu}=\hat{x}_{\mu} 2\pi/L_{\mu}$ with
$\hat{x}_{\mu}$ being a unit vector along $\mu$-direction. We further
define the following product on a plaquette as shown in
Fig.~\ref{fig:plaquette},
\begin{subequations} %13:50'ÌŽ®ŒQ
 \label{field_stlength}
\begin{align}
&Z(\bm{k})
 =V_{\bm{k},\bm{k}+\hat{\bm{k}}_{1}}
  V_{\bm{k}+\hat{\bm{k}}_{1},\bm{k}+\hat{\bm{k}}_{1}+\hat{\bm{k}}_{2}}
  V_{\bm{k}+\hat{\bm{k}}_{1}+\hat{\bm{k}}_{2},\bm{k}+\hat{\bm{k}}_{2}}
  V_{\bm{k}+\hat{\bm{k}}_{2},\bm{k}},
 \label{field_stlength_Z}\\
&F_{12}(\bm{k})=\ln\,Z(\bm{k}),
 \label{field_stlength_F_12}
\end{align}
\end{subequations}
where we determine the principal branch by restricting the region of
$F_{12}(\bm{k})$ as
\begin{equation}
 -\pi<-\i\,F_{12}(\bm{k})\leq\pi.
\end{equation}
Note that \eqref{U(1)_link} and \eqref{field_stlength_F_12} are related
to the link variable and the Wilson loop, respectively in the lattice
gauge theory.  Then the Chern number is give as
\begin{equation}
 \nu=-\frac{1}{2\pi\i}\sum_{\bm{k}}F_{12}(\bm{k}).
  \label{FHS_chern}
\end{equation}
The continuum representation is given in Appendix~\ref{sec:Berry_con}.

%%%%%%%%%%%%%%%%%%%%%%%%%%%%%%%%%%%%%%%%%%%%%%%%%%%%%%%%%%%%%%%%%%%%%%%
%% Coh-Vanderbilt's method
%%%%%%%%%%%%%%%%%%%%%%%%%%%%%%%%%%%%%%%%%%%%%%%%%%%%%%%%%%%%%%%%%%%%%%%
\subsection{Coh-Vanderbilt's method}
On the other hand, Coh and Vanderbilt related the Chern number and the
electronic polarization by flux insertion \cite{Coh-V}. Later, this
method has been discussed in details by Kang, Lee, and Cho
\cite{Kang-L-C}. In this method, the Chern number is given by the
following relations,
\begin{subequations} %11:31'ÌŽ®ŒQ
 \label{CV_method}
\begin{align}
 z(\phi)
 &=\braket{\Psi_{0}(0,\phi)|U_{1}|\Psi_{0}(0,\phi)}
 \label{overlap_flux0}\\
 &=-\prod_{\bm{k}}
 V_{\bm{k}+\frac{\phi}{2\pi}\hat{\bm{k}}_2,
 \bm{k}+\frac{\phi}{2\pi}\hat{\bm{k}}_2-\hat{\bm{k}}_1},
 \label{overlap_flux1}\\
 \nu&=-\frac{\partial}{\partial\phi}\mathrm{Im\,ln\,}z(\phi)
 \label{chern_CV}
\end{align}
\end{subequations}
where the polarization operator along $\mu$ direction is
\begin{equation}
 U_{\mu}=\exp\left(\,\i\frac{2\pi}{L_{\mu}}
	      \sum_{\bm{r}}x_{\mu}\,n_{\bm{r}}\right).
 \label{LSMO.23}
\end{equation}
This is the twist operator introduced in the LSM theorem for $d\geq 2$
\cite{Oshikawa-2000a,Oshikawa-2000b}.
$\ket{\Psi_{0}(\phi_{1},\phi_{2})}$ is the many-body ground state as a
function of the flux $\phi_{\mu}$. The Chern number $\nu$ is quantized
in the thermodynamic limit.

As discussed in Appendix~\ref{sec:Berry_con}, the Chern number in the
continuum limit is given by different gauges as that of FHS's
method. Wheres the gauge of FHS's method is like the symmetric gauge,
that of CV's method corresponds to the Landau gauge.  However, we can
modify the CV's method so that the Chern number is given by symmetric
gauge. For example, we define the polarization as follows,
\begin{align}
 z(\phi)
 &=\braket{\Psi_{0}(\phi,\phi)|
 U_1^{\mathstrut}U_{2}^{-1}|\Psi_{0}(\phi,\phi)}\\
 &=-\prod_{\bm{k}}
 V_{\bm{k}+\frac{\phi}{2\pi}(\hat{\bm{k}}_1+\hat{\bm{k}}_2),
 \bm{k}+\frac{\phi}{2\pi}(\hat{\bm{k}}_1+\hat{\bm{k}}_2)
 -\hat{\bm{k}}_1+\hat{\bm{k}}_2}.
 \label{flux_chern_number2}
\end{align}
This gives the same representation for $L_1=L_2$ as that of FHS's method
in the continuum limit.

%%%%%%%%%%%%%%%%%%%%%%%%%%%%%%%%%%%%%%%%%%%%%%%%%%%%%%%%%%%%%%%%%%%%%%%
%% Spiral Boundary Conditions
%%%%%%%%%%%%%%%%%%%%%%%%%%%%%%%%%%%%%%%%%%%%%%%%%%%%%%%%%%%%%%%%%%%%%%%
\section{Spiral Boundary Conditions}
\label{sec:SBC}

For 2D systems, the exponential position operator with SBCs is given by
\begin{equation}
 U_{\rm SBC}=\exp\left(\i\sum_{j=1}^L\frac{2\pi jn_{j}}{L}\right),
 \qquad L=L_1L_2,
\end{equation}
where $j$ is the site number of the extended 1D chain which sweeps all
lattice sites in 1D order as shown in Fig.\ref{fig:unit_SBC}(b). On the
other hand, for conventional PBCs as in Fig.\ref{fig:unit_SBC}(a), there
are two operators as Eq.~(\ref{LSMO.23}).  In this section, we discuss
the relationship between $U_{\rm SBC}$ and $U_{\mu}$ for 2D, and how the
wave number $\bm{k}$ appearing in Sec.~\ref{sec:Chern_number} should be
replaced for SBCs.

\subsection{Definition of spiral boundary conditions}

In order to consider the flux insertion to the system with SBCs, we
reinterpret the twist operators in the Oshikawa's argument
(\ref{LSMO.23}). First, we consider the unit vectors in the real space
and the inverse lattice space of 2D systems as
\begin{subequations}
\begin{alignat}{3}
 \bm{A}_1&=(L_1,0), &\qquad \bm{B}_1&=2\pi(\tfrac{1}{L_1},0),
 \label{AB1}\\
 \bm{A}_2&=(0,L_2), &  \bm{B}_2&=2\pi(0,\tfrac{1}{L_2}).
 \label{AB2}
\end{alignat}
\end{subequations}
Then these vectors satisfy the following relation,
\begin{equation}
 \bm{A}_{\mu}\cdot\bm{B}_{\nu}=2\pi\delta_{\mu,\nu}.
  \label{defAB}
\end{equation}
The twist operator (\ref{LSMO.23}) is written using the above unit
vectors in the inverse lattice space as
\begin{equation}
 U_{\mu}
  =\exp\left(\i\sum_{\bm{r}}\bm{B}_{\mu}\cdot\bm{r}n_{\bm{r}}\right).
  \label{LSMYO.21}
\end{equation}

Next, we choose the unit vectors for SBCs as shown in
Fig.~\ref{fig:unit_SBC}(b) so that they satisfy the relation
(\ref{defAB}).  Then we have
\begin{subequations}
 \label{A'B'}
\begin{alignat}{3}
 \bm{A}'_1&=(L_1,-1),&\qquad \bm{B}'_1&=2\pi(\tfrac{1}{L_1},0),
  \label{AB1'}\\
 \bm{A}'_2&=(0,L_2),&
 \bm{B}'_2&=2\pi(\tfrac{1}{L_1L_2},\tfrac{1}{L_2}).
 \label{AB2'}
\end{alignat}
\end{subequations}
The twist operators in 2D systems with SBCs are introduced as
\begin{equation}
 U'_{\mu}
  =\exp\left(\i\sum_{\bm{r}}\bm{B}'_{\mu}\cdot\bm{r}n_{\bm{r}}\right).
  \label{LSMYO.22}
\end{equation}
For $\mu=1$, due to the relation $\bm{B}_1'=\bm{B}_1$, $U_1'$ is the
same operator as $U_1$ with SBCs. In order to represent $U_1$ in SBCs,
we use the following relation
\begin{equation}
 \frac{2\pi jn_{j}}{L_1}=\frac{2\pi x_1n_{\bm{r}}}{L_1}
  \quad (\mathrm{mod}\,2\pi).
  \label{U1mod}
\end{equation}
For $\mu=2$, the exponent part of the twist operator is confirmed as
that of SBCs as
\begin{align}
 \sum_{\bm{r}}\bm{B}'_2\cdot\bm{r}n_{\bm{r}}
  &=\sum_{i_1=1}^{L_1}\sum_{i_2=0}^{L_2-1}
  2\pi\frac{i_1+i_2L_1}{L_1L_2}n_{(i_1,i_2)}\\
 &=\sum_{j=1}^{L}\frac{2\pi}{L}jn_{j},
\end{align}
where $L=L_1L_2$ and $j=i_1+i_2L_1$.  Thus the twist operators for SBCs
are identified as
\begin{subequations}
  \label{correspondence_A_for_U}
\begin{align}
 U_1'&=\exp\left(\i\sum_{j=1}^L\frac{2\pi jn_{j}}{L_1}\right)
 =(U_{\rm SBC})^{L_2},\label{correspondence_A_for_U1}\\
 U_2'&=U_{\rm SBC}.\label{correspondence_A_for_U2}
\end{align}
\end{subequations}
These correspondences mean that the flux insertion for the extended 1D
chain of SBCs gives rise an effective current along $y$-direction, while
$U_1'$ causes a current only to $x$-direction due to the relation
(\ref{U1mod}).

In the same way, the unit vectors of the system with SBCs for
$(\pi,\pi)$ order illustrated in Fig.~\ref{fig:unit_SBC}(c), are
identified as
\begin{subequations}
\begin{alignat}{3}
 \bm{A}''_1&=(L_1-1,-1),&~&
 \bm{B}''_1&=\tfrac{2\pi}{L_1L_2+L_1-L_2}(L_2,-L_1),\\
 \bm{A}''_2&=(L_1,L_2),& &
 \bm{B}''_2&=\tfrac{2\pi}{L_1L_2+L_1-L_2}(1,L_1-1).
\end{alignat}
\end{subequations}
Then, the exponential position operators in these SBCs $U_j''$ for the
isotropic case $L_1=L_2$ are related as
\begin{equation}
 U_1''=U_1^{\mathstrut}U_2^{-1},\qquad U_2''=U_{\rm SBC}.
  \label{correspondence_B_for_U}
\end{equation}
Since we are not interested in charge orders in this paper, we will not
use these boundary conditions.

\subsection{Redefinition of Coh-Vanderbilt's method}

In the Coh-Vanderbilt method, we need to apply flux to the system to
generate a current in the vertical direction of the twist operator for
calculating the Chern number.  However, since the flux is continuum
number, we can not distinguish whether the direction of the flux is
along $\bm{B}_1'$ or $\bm{B}_2'$. On the other hand, roles of twist
operators for each directions are apparently distinguished as in
Eq.~(\ref{correspondence_A_for_U}). Therefore, we apply the flux to
$\bm{B}_2'$ direction which is along the extended 1D chain, and turn our
attention to the response to $\bm{B}_1'$ direction.

It follows from Eq.~(\ref{correspondence_A_for_U}) that the expectation
value of the twist operator with flux (\ref{overlap_flux0}) for the
system with SBCs is redefined as
\begin{equation}
 z'(\phi)=\braket{\Psi_{0}'(\phi)|U_1'|\Psi_{0}'(\phi)},
\end{equation}
where $\ket{\Psi_{0}'(\phi)}$ is the many-body wave function of the
ground state with flux $\phi$.  The flux $\phi$ is imposed to the
extended 1D chain with $L=L_1L_2$ sites. From Eq.~(\ref{CV_method}), the
following relation is satisfied,
\begin{equation}
 z'(\phi)=\e^{-\i\nu\phi}z'(0).
  \label{z'(phi)}
\end{equation}
If we consider the expectation value of $U_2'$ as
\begin{equation}
 z_2'(\phi)=\braket{\Psi_{0}'(\phi)|U_2'|\Psi_{0}'(\phi)},
  \label{z2'(phi)}
\end{equation}
we can not detect the Chern number, because the applied current by the
flux and its response are in the parallel directions and there is no
perpendicular component in the response.

\subsection{Correspondence of wave numbers}

Now we discuss the correspondence of the wave numbers defined in 2D PBCs
with those of SBCs.  Let us consider a tight-biding model defined on a
2D square lattice. When we apply SBCs to this system, the hopping term
is written as
\begin{align}
 \mathcal{H}
  =&-t
 \sum_{i=1}^{L}
 \left[(c_i^{\dag}c_{i+1}^{\mathstrut}+\mbox{H.c.})
  +(c_i^{\dag}c_{i+\Lambda}^{\mathstrut}+\mbox{H.c.})
 \right]\nonumber\\
 =&-2t\sum_k
 \left[\cos k+\cos\Lambda k\right]c_k^{\dag}c_{k}^{\mathstrut},
\end{align}
where $i$ is the number of the lattice sites of the extended 1D chain
with $L\equiv L_1L_2$ sites, and $\Lambda$ is a parameter depending on
the number of lattice sites in $x$-direction.  As shown in
Fig.~\ref{fig:unit_SBC}, this parameter is related to ways to label the
sites of extended 1D chain and $\Lambda\in\mathbb{Z}$. For SBCs for
$(\pi,0)$ order $\Lambda=L_1$ while those for $(\pi,\pi)$ order,
$\Lambda=L_1-1$ .  In the present case, we are not interested in charge
orders, so that we choose SBCs for $(\pi,0)$ order for simplicity.  The
Fourier transformation of the extended 1D chain is given by
\begin{equation}
 c_{k,\alpha}=\frac{1}{\sqrt{L}}\sum_{j=1}^L\e^{-\i k x_j}c_{j,\alpha}.
\end{equation}
Then the 2D wave vectors are replaced as the 1D wave number by
\begin{subequations} %17:01'ÌŽ®ŒQ
\label{replacement-0}
\begin{gather}
 \bm{k}=(k_{x},k_{y})\to (k,L_1 k),\label{replacement-0a}\\
 k=\frac{2\pi}{L}n,\qquad n=0,1,2,\cdots,L,
 \label{replacement-0b}
\end{gather}
\end{subequations}

Next we consider the unit wave numbers $\hat{\bm{k}}_{\mu}$ ($\mu=1,2$)
appearing in Sec.~\ref{sec:Chern_number}. According to the results of
$\bm{B}_{\mu}'$ given in Eq.~(\ref{A'B'}), there are following
relations,
\begin{subequations}
\label{replacement-1}
\begin{align}
 \hat{\bm{k}}_1&=(\hat{k}_1,0)\to(L_2\hat{k},0),
 \label{replacement-1a}\\
  \hat{\bm{k}}_2&=(0,\hat{k}_2)\to(\hat{k},L_1 \hat{k}),
 \label{replacement-1b}
\end{align}
\end{subequations}
where
\begin{equation}
 \hat{k}_{\mu}=\frac{2\pi}{L_{\mu}},\qquad \hat{k}=\frac{2\pi}{L}.
\end{equation}
Then $Z(\bm{k})$ defined by Eq.~(\ref{field_stlength_Z}) and $z$ by
Eq.~(\ref{overlap_flux1}) are replaced as
\begin{align}
&Z(\bm{k})
 \to V_{(k,L_1 k),(k+L_2\hat{k},L_1 k)}
 V_{(k+L_2\hat{k},L_1 k),
 (k+L_2\hat{k}+\hat{k},L_1 k+L_1\hat{k})}\nonumber\\
 &\times V_{(k+L_2\hat{k}+\hat{k},L_1 k+L_1\hat{k}),
 (k+\hat{k},L_1 k+L_1\hat{k})}
 V_{(k+\hat{k},L_1 k+L_1\hat{k}),(k,L_1 k)},\\
 &z'(\phi)\to
 -\prod_{k}V_{(k+\tfrac{\phi}{L},L_1 k+\tfrac{\phi}{L_2}),
 (k+\tfrac{\phi}{L}-L_2\hat{k},L_1 k+\tfrac{\phi}{L_2})}.
\end{align}

%%%%%%%%%%%%%%%%%%%%%%%%%%%%%%%%%%%%%%%%%%%%%%%%%%%%%%%%%%%%%%%%%%%%%%%
%% Results
%%%%%%%%%%%%%%%%%%%%%%%%%%%%%%%%%%%%%%%%%%%%%%%%%%%%%%%%%%%%%%%%%%%%%%%

\begin{figure}[t]
 \includegraphics[width=7.0cm]{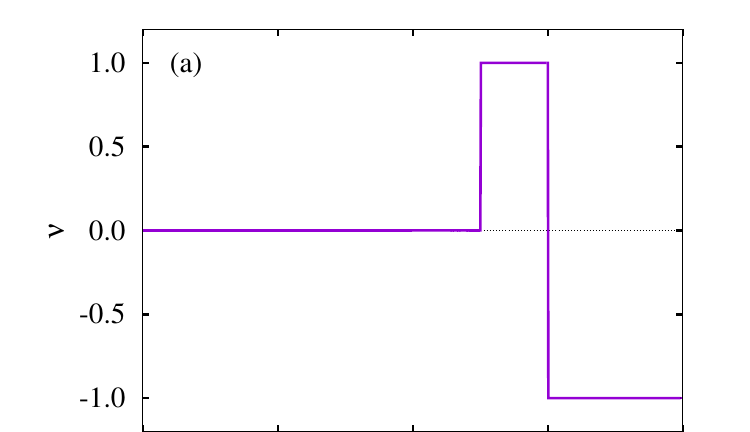}
 \includegraphics[width=7.0cm]{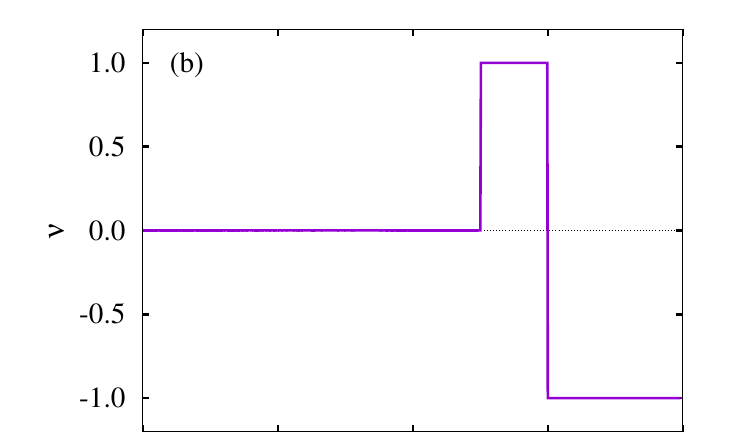}
 \includegraphics[width=7.0cm]{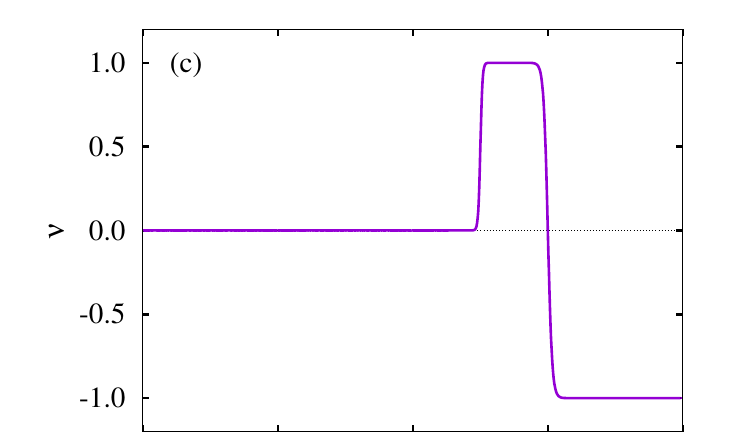}
 \includegraphics[width=7.0cm]{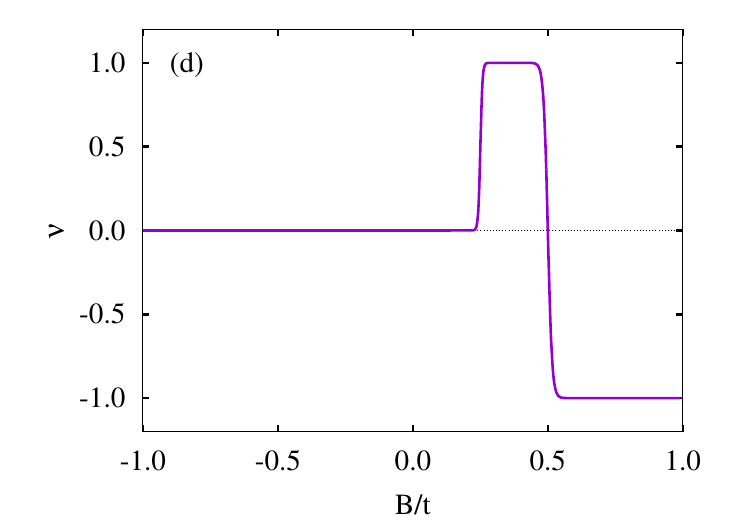}
\caption{Chern numbers of the 2D WD model with $M/t=1$ and $L=64\times
 64$ sites calculated by Fukui-Hatsugai-Suzuki's method for (a)
 conventional PBCs and (b) SBCs, and those calculated by
 Coh-Vanderbilt's method for (c) conventional PBCs and (d)
 SBCs. Topological regions are $1/4\leq B/t\leq 1/2$ with $\nu=1$ and
 $1/2\leq B/t$ with $\nu=-1$.}
\label{fig:chern_WD}
\end{figure}

\section{Results}
\label{sec:Results}
We demonstrate the above discussions in representative models for Chern
insulators: the Wilson-Dirac model and the Haldane model.

\subsection{Wilson-Dirac model}

As a fundamental model to describe 2D Chern insulators, we consider the
Wilson-Dirac model \cite{Wilson,Qi-W-Z},
\begin{align}
 \mathcal{H}=&\frac{-\i t}{2}\sum_{j,\mu=x,y}(c_j^{\dag}
 \tau_{\mu}c_{j+\hat{\mu}}^{\mathstrut}-\mathrm{H.c.})
 +(M-B)\sum_jc_j^{\dag}\tau_{3}c_{j}^{\mathstrut}\nonumber\\
 &+\frac{B}{2}\sum_{j,\mu=x,y}
 (c_j^{\dag}\tau_{3}c_{j+\hat{\mu}}^{\mathstrut}+\mathrm{H.c.}).
\end{align}
The Fourier representation of this model becomes
\begin{subequations} %00:41'ÌŽ®ŒQ
 \label{WD}
\begin{align}
 \mathcal{H}
  =&\sum_{\bm{k},\alpha, \beta} c^\dag_{\bm{k}, \alpha}
  \,H_{\alpha \beta}(\bm{k}) \,c^{\mathstrut}_{\bm{k}, \beta},
 \label{WD.1}\\
  H(\bm{k})
  =&t\sum_{\mu =x, y}\sin k_\mu \,\tau_\mu
 +\left[M - B \sum_{\mu = x, y}
 \bigl(1 -\cos k_\mu \bigr)\right]\tau_z,
 \label{WD.2}
\end{align}
\end{subequations}
where $t$ is the hopping amplitude, $M$ is the mass, $B$ is the
coefficient of the Wilson term, $c_{\bm{k},\alpha}$ is the annihilation
operator of a fermion with a 2D wave number, $\alpha,\beta$ are orbital
indices, and $\tau_{\mu}$ are the Pauli matrices. The energy eigenvalue
is given by
\begin{equation}
 \varepsilon_{\bm{k}}^2=t^2(\sin^2 k_x+\sin^2 k_y)
  +\bigl\{M - B(2-\cos k_x -\cos k_y)\bigr\}^2.
  \label{sxy_continuum}
\end{equation}
This system is a trivial insulator for $B/t<M/4t$, and a topological
insulator with $\nu=1$ for $M/4t\leq B/t\leq M/2t$ and that with
$\nu=-1$ for $M/2t\leq B/t$.  Topological phase transition occurs at
$B=M/4$ and $M/2$ \cite{Imura-Y-M-H-K}.  These transition points can be
identified by vanishing of the bulk energy gap $\varepsilon_{\bm{k}}=0$.
In the continuum version of the model, the Hall conductivity is
calculated as \cite{So}
\begin{equation}
  \sigma_{xy}=-\frac{e^2}{2h}[\sgn(M)+\sgn(B)].
\end{equation}
Therefore the system is a topological (trivial) insulator for $MB>0$
($MB<0$), and a topological transition occurs at $B=0$ for fixed $M$.

\begin{figure}[t]
 \includegraphics[width=7.0cm]{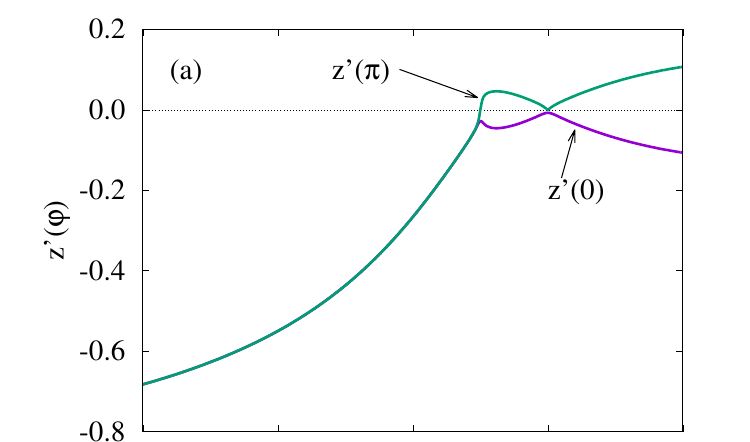}
 \includegraphics[width=7.0cm]{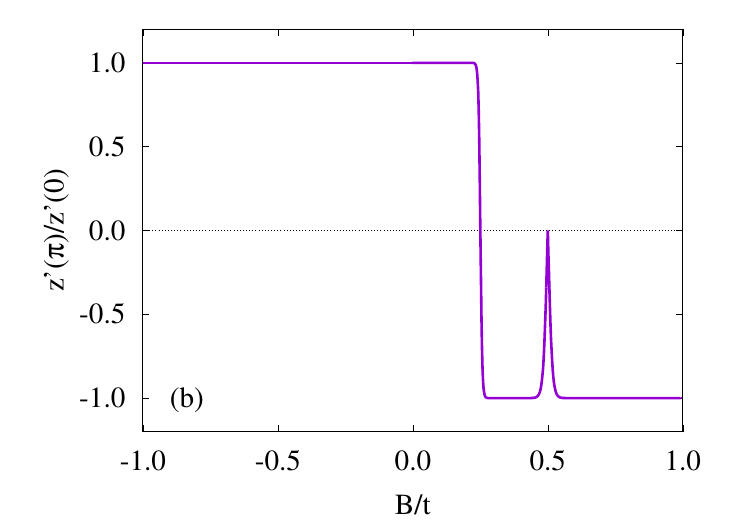}
\caption{(a) Expectation values of the twist operator $U_1'$ with and
 without flux, $z'(0)$ and $z'(\pi)$. (b) The ratio of the expectation
 values $z'(\pi)/z'(0)$. There is no sign change for $z'(0)$, while
 $z'(\pi)$ has the opposite sign in the topological regions. These
 results are calculated for the systems with $M/t=1$ and $L=64\times 64$
 sites.}
\label{fig:z_WD_flux}
\end{figure}

% result-WD

As shown in Fig.~\ref{fig:chern_WD}, the Chern number of the WD model
obtained by the FHS method and the CV method with conventional PBCs and
SBCs. The finite-size effect of the FHS method tends to be smaller than
that of the CV method, and that of the conventional PBCs tends to be
smaller than the SBCs.

% z'()
Figure~\ref{fig:z_WD_flux}(a) shows $z'(\phi)$, the ground-state
expectation values of the twist operator $U_1'$ in SBCs. According to
the results, $z'(0)$ does not change the sign. On the other hand,
$z'(\pi)$ changes the sign in the topological regions $M/4<B$ with the
Chern number $\nu=\pm 1$, due to the relation (\ref{z'(phi)}).  We can
also identify the topological regions by calculating the ratio of the
expectation values of $U_1'$ with and without flux $z'(\pi)/z'(0)$ as
shown in Fig.~\ref{fig:z_WD_flux}(b).

In case we consider the expectation value of $U_{\rm SBC}$ as
(\ref{z2'(phi)}), it has the opposite sign in the second topological
region $M/2<B$, and that with different SBCs shown in
Fig.~\ref{fig:unit_SBC}(c) has opposite sign in the both two topological
regions $M/4<B$.  These results are already discussed in
Ref.~\onlinecite{Nakamura-M-N}.

\subsection{Haldane model}

\begin{figure}[t]
\begin{center}
 \includegraphics[width=9cm]{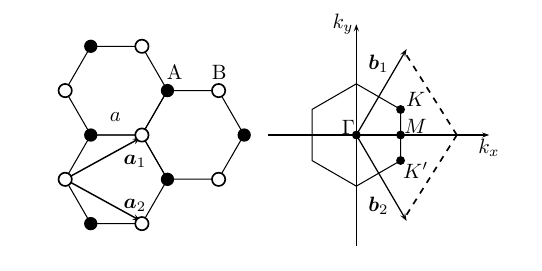}
\end{center}
 \caption{Unit vectors and reciprocal lattice vectors of the honeycomb
 lattice.}
 \label{fig:honeycomb}
\end{figure}

As another fundamental model for a Chern insulator, we consider the
Haldane model~\cite{Haldane1988} describing the fermions on a honeycomb
lattice with an alternating potential $M$ and a next-nearest
neighbor hopping $\kappa$,
\begin{equation}
 \mathcal{H}=t\sum_{\braket{ij}} c_i^\dagger c_j^{\mathstrut}
  + M\sum_i  \eta_i c_i^\dagger c_i^{\mathstrut}
  + \i\frac{\kappa}{3\sqrt{3}} \sum_{\braket{\braket{ij}}}
  \nu_{ij} c_i^\dagger c_j^{\mathstrut},
\end{equation}
where $c_i^{\dag}$ ($c_i$) is a creation (annihilation) operator at site
$i$ (spin indices are omitted).  $\braket{ij}$ and
$\braket{\braket{ij}}$, denote a nearest and a next nearest pair,
respectively. $\eta_i=1$ ($\eta_i=-1$) for A (B) sublattice, and
$\nu_{ij}=(2/\sqrt{3})(\hat{\bm{d}}_1\times\hat{\bm{d}}_2)_z=\pm 1$,
where $\hat{\bm{d}}_1$ and $\hat{\bm{d}}_2$ are unit vectors along the
two bonds on which the electron hops from a site $j$ to $i$, as shown in
Fig.~\ref{fig:honeycomb}.

For the honeycomb lattice, the fundamental vectors are
\begin{equation}
 \bm{a}_{1}  = \frac{a}{2}(3, \sqrt{3}),\qquad
  \bm{a}_{2}  = \frac{a}{2}(3, -\sqrt{3}).
\end{equation}
Then the reciprocal lattice vectors are chosen as
\begin{equation}
 \bm{b}_{1} = \frac{2\pi}{3a}(1, \sqrt{3}),\qquad
  \bm{b}_{2} = \frac{2\pi}{3a}(1, -\sqrt{3}),
\end{equation}
by the following relation,
\begin{equation}
 \bm{a}_i\cdot\bm{b}_j=2\pi\delta_{ij}.
\end{equation}
The wave numbers are given in these bases as
\begin{equation}
 \bm{k}=k_1\bm{b}_{1}+k_2\bm{b}_{2},
\end{equation}
where $k_i\in [-\pi,\pi]$. 

\begin{figure}[t]
 \includegraphics[width=7.0cm]{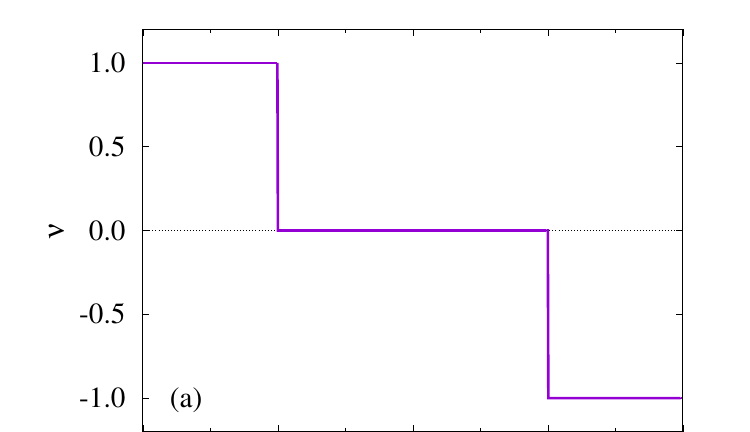}
 \includegraphics[width=7.0cm]{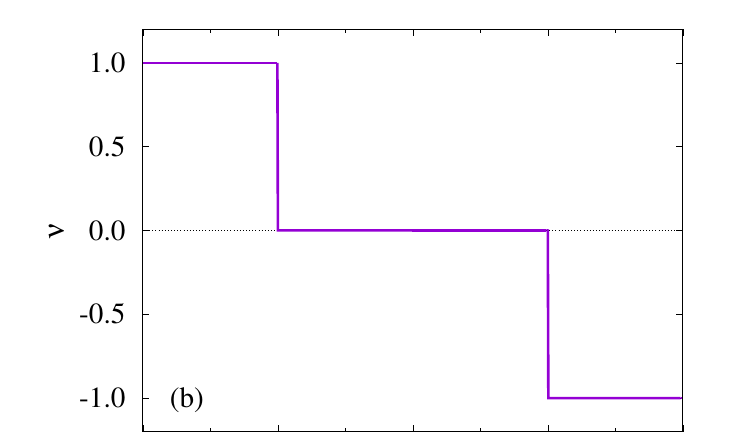}
 \includegraphics[width=7.0cm]{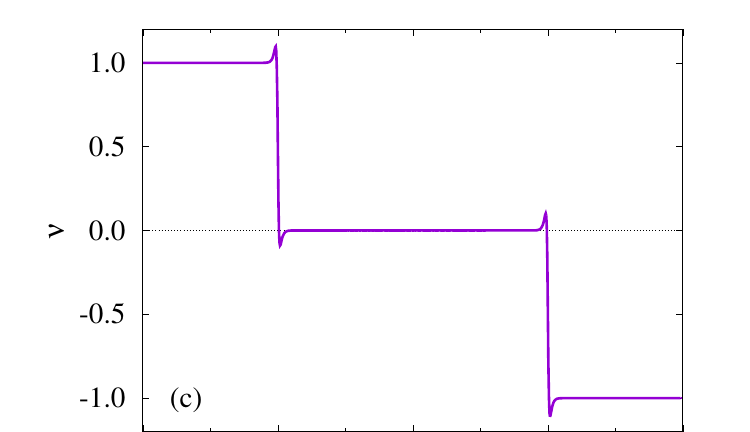}
 \includegraphics[width=7.0cm]{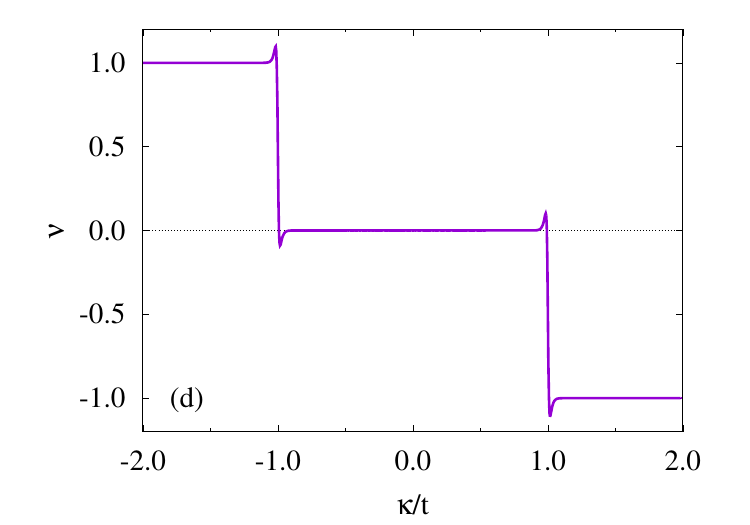}
\caption{Chern numbers of the Haldane model with $M/t=1$ and $L=64\times
 64$ sites calculated by Fukui-Hatsugai-Suzuki's method for (a)
 conventional PBCs and (b) SBCs, and those calculated by
 Coh-Vanderbilt's method for (c) conventional PBCs and (d)
 SBCs. Topological regions are $\kappa/t\leq -1$ with $\nu=1$ and $1
 \leq \kappa/t$ with $\nu=-1$.}
\label{fig:chern_Hal}
\end{figure}

The Hamiltonian in the momentum space has the following form,
\begin{equation}
% H(\bm{k})=\varepsilon(\bm{k})\tau_0+d_i(\bm{k})\tau_i,
 H(\bm{k})=d_i(\bm{k})\tau_i,
\end{equation}
with the energy
\begin{equation}
% E(\bm{k})=\varepsilon(\bm{k})\pm\sqrt{\sum_id^2_i(\bm{k})}.
 \varepsilon(\bm{k})=\pm\sqrt{\sum_id^2_i(\bm{k})}.
\end{equation}
The parameters in Hamiltonian are given as follows,
\begin{subequations}
\begin{align}
 d_1&=t[\cos(\bm{k}\cdot\bm{a}_1)+\cos(\bm{k}\cdot\bm{a}_2)+1],\\
 d_2&=t[\sin(\bm{k}\cdot\bm{a}_1)+\sin(\bm{k}\cdot\bm{a}_2)],\\
 d_3&=M+\frac{2\kappa}{3\sqrt{3}}\nonumber\\
%\sin\phi\\
 \times&
 [\sin(\bm{k}\cdot\bm{a}_1)-\sin(\bm{k}\cdot\bm{a}_2)
 -\sin(\bm{k}\cdot(\bm{a}_1-\bm{a}_2))].
%,\nonumber\\
% \varepsilon(\bm{k})&=0.
% \varepsilon(\bm{k})&=\frac{2\kappa}{3\sqrt{3}}\\
% \times&
% [\cos(\bm{k}\cdot\bm{a}_1)+\cos(\bm{k}\cdot\bm{a}_2)
% +\cos(\bm{k}\cdot(\bm{a}_1-\bm{a}_2))].\nonumber
\end{align}
\end{subequations}
Thus we can calculate the Chern number as in the same way of the square
lattice systems, by using the wave vector as $\bm{k}=(k_1,k_2)$.

Then the Hall conductivity of the system at the charge neutrality point
becomes
\begin{equation}
 \sigma_{xy}=\frac{e^2}{h}
  \sgn(\kappa)\theta(|\kappa|-|M|).
\end{equation}
This means that the system is a Chern insulator for $|\kappa|>|M|$ with
$\nu=\pm 1$ and a trivial insulator for $|\kappa|<|M|$ with $\nu=0$.

% result-Haldane

Figure~\ref{fig:chern_Hal} shows the Chern number of the Haldane model
is obtained by the FHS method and the CV method with conventional PBCs
and SBCs. Finite-size effect of the FHS method tends to be smaller than
that of the CV method, and that of the conventional PBCs tends to be
smaller than the SBCs. These tendencies are same for the WD model.

\begin{figure}[t]
 \includegraphics[width=7.0cm]{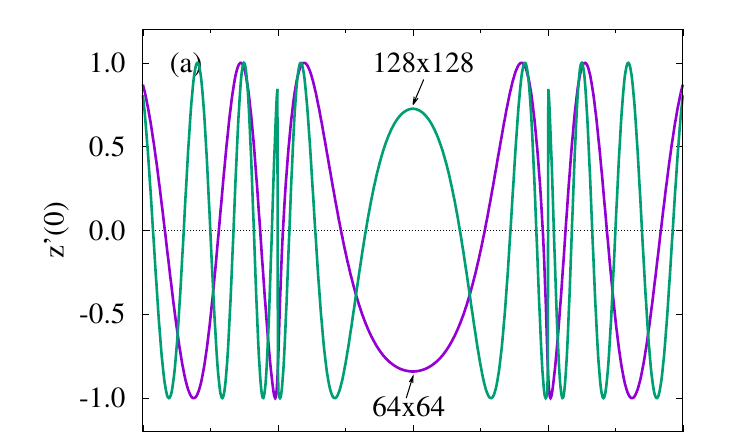}
 \includegraphics[width=7.0cm]{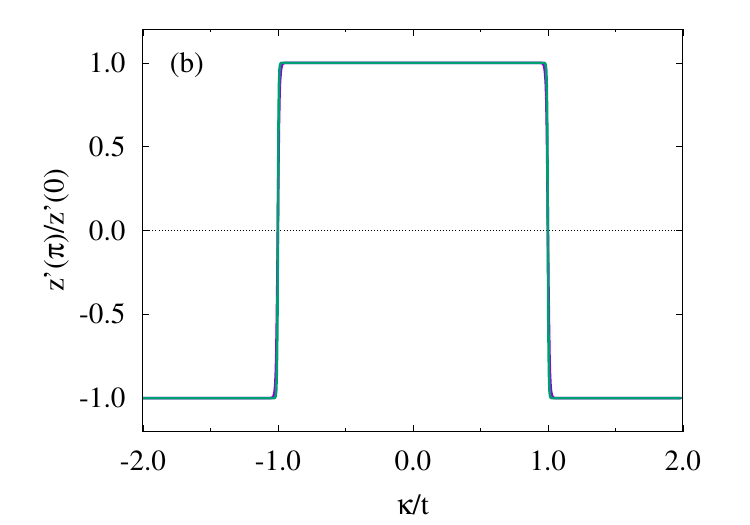}
\caption{(a) Expectation values of twist operators $U_1'$ without flux
 $z'(0)$. (b) The ratio of the expectation values of $U_1'$ with and
 without flux $z'(\pi)/z'(0)$ calculated in the systems with $L=64\times
 64$ and $L=128\times 128$.}
\label{fig:z_Hal_flux}
\end{figure}

% z'()
Figure~\ref{fig:z_Hal_flux}(a) shows the ground-state expectation values
of the twist operators in SBCs $z'(\phi)$ without flux. The result of
$z'(0)$ has large finite-size effects with oscillations, so that it is
difficult to characterize the insulating states and electronic
polarization. This behavior is considered to be an effect of the next
nearest neighbor hopping process. However, the ratio of $z'(\phi)$ with
and without flux $z'(\pi)/z'(0)$, this well characterizes the
topological regions $|\kappa|<|M|$ as shown in
Fig.~\ref{fig:z_Hal_flux}(b).

%%%%%%%%%%%%%%%%%%%%%%%%%%%%%%%%%%%%%%%%%%%%%%%%%%%%%%%%%%%%%%%%%%%%%%%
%% Summary and discussion
%%%%%%%%%%%%%%%%%%%%%%%%%%%%%%%%%%%%%%%%%%%%%%%%%%%%%%%%%%%%%%%%%%%%%%%
\section{Summary and discussion}
\label{sec:Summary}

In summary, we have explored the relationship between electronic
polarization and the Chern number in 2D systems with SBCs. Initially, we
examined two methods for calculating Chern numbers in 2D lattice
systems: Fukui-Hatsugai-Suzuki's method and Coh-Vanderbilt's
method. Subsequently, we introduced twist operators in 2D systems with
SBCs and redefined the aforementioned methods using the wave numbers of
the extended 1D chain. The crucial aspect of this discussion is that
flux insertion into the extended 1D chain generates an effective current
along the $y$-direction, and the twist operator detects the response to
the $x$-direction. Finally, we illustrated the above discussions in
representative models for Chern insulators, such as the Wilson-Dirac
model and the Haldane model, demonstrating that the calculation of Chern
numbers and the detection of topological phases are achievable using
methods with SBCs.

% physical picture

The relationship between topological states in 1D and those in 2D with a
Chern number is considered as follows: As mentioned in
Sec.~\ref{sec:intro}, the signs of $z$ classify 1D gapped states, and in
some cases, such as the Su-Schrieffer-Heeger model
\cite{Su-S-H1979,Su-S-H1980}, an insulating state with localized edge
electrons emerges. If we then consider creating a 2D system by
connecting these 1D edge states with spiral boundary conditions (SBCs),
an insulating state with edge current may appear in response to applied
flux. This state represents a 2D topological state with a Chern number.

% general dimension

As an extension of the argument presented in this paper, we may consider
SBCs in general dimensions. According to the discussion in
Sec.~\ref{sec:SBC}, the twist operators in $d$-dimensional systems are
generalized as follows:
\begin{equation}
 U'_{\mu}
  =\exp\left(\i\sum_{j=1}^L\frac{2\pi
	jn_{j}}{\prod_{k=1}^{\mu}L_k}\right),
  \quad L=\prod_{k=1}^{d}L_k.
\end{equation}
By utilizing these operators and incorporating flux insertions, we may
identify $d$-dimensional topological phases and higher-order topological
states.

The present method with SBCs is expected to be applicable to numerical
calculations, such as the density matrix renormalization group
method. For systems in more than 2D, finite-size scaling is not as
straightforward as in 1D systems because increasing the number of
lattice sites cannot be done linearly while maintaining the symmetry of
the unit systems. However, in systems with SBCs, we can increase the
lattice size linearly, enabling finite-size scaling similar to 1D
systems \cite{Kadosawa-N-O-N1,Kadosawa-N-O-N2,Kadosawa-N-O-N3}. We
anticipate that the present method will prove useful for the numerical
analysis of topological systems with electron-electron interactions.

%%%%%%%%%%%%%%%%%%%%%%%%%%%%%%%%%%%%%%%%%%%%%%%%%%%%%%%%%%%%%%%%%%%%%%%
%% Acknowledgment
%%%%%%%%%%%%%%%%%%%%%%%%%%%%%%%%%%%%%%%%%%%%%%%%%%%%%%%%%%%%%%%%%%%%%%%
\section{Acknowledgments}
M. N. acknowledges the research fellow position of the Institute of
Industrial Science, The University of Tokyo.  M.~N. is supported partly
by MEXT/JSPS KAKENHI Grant Number JP20K03769.  The authors thank
S. Nishimoto and H. Watanabe for helpful discussions.

\appendix

\section{Continuum limit}
\label{sec:Berry_con}

%In this section, we discuss how the methods by Fukui-Hatsugai-Suzuki and
%by Coh-Vanderbilt for lattice systems give the Chern number, considering
%the continuum limits.

In this section, we discuss the continuum limits of the formulas for the
Chern number given by the Fukui-Hatsugai-Suzuki's method and by the
Coh-Vanderbilt's method for lattice systems, to compare the
Thouless-Kohmoto-Nightingale-den Nijs (TKNN) formula.

\subsection{TKNN formula}

Let's consider a 2D system.  When the system is translationally
invariant, the Hamiltonian satisfies the following eigenvalue equation
in term of the Bloch state,
\begin{equation}
 H(\bm{k})\ket{u_{n}(\bm{k})}=E_{n}(\bm{k})\ket{u_{n}(\bm{k})},
\end{equation}
where $\ket{u_{n}(\bm{k})}$ is the Bloch eigenstate of the $n$-th band,
and normalized as $\braket{u_{n}(\bm{k})|u_{m}(\bm{k})}=\delta_{n,m}$.
The Berry connection of the $n$-th band is defined as
\begin{equation}
 \bm{A}^{(n)}(\bm{k})
  =\i\braket{u_{n}(\bm{k})|\nabla_{\bm{k}}u_{n}(\bm{k})}.
  \label{Berry_connection}
\end{equation}
Then the Berry curvature is given as
\begin{equation}
 \bm{B}^{(n)}(\bm{k})=\nabla_{\bm{k}}\times \bm{A}^{(n)}(\bm{k}).
\label{Berry_curvature}
\end{equation}
The Berry connection and the Berry curvature correspond to the vector
potential and the magnetic field in the electrodynamics, respectively.

The Chern number for the $n$-th band $\nu_{n}$ is given by the Berry
curvature as follows,
\begin{align}
 \nu_{n}
 &=\frac{1}{2\pi}\iint_{\rm BZ}\d^2 \bm{k} \,[\bm{B}^{(n)}(\bm{k})]_{z}
 \nonumber\\
 &=\frac{1}{2\pi}\oint_{\partial{\rm BZ}}\d \bm{k}\cdot\bm{A}^{(n)}(\bm{k}),
 \label{chern_number_continuum}
\end{align}
where BZ means the 1st Brillouin zone.  Then the Chern number $\nu$ is
given by the summation of $\nu_{n}$ over the occupied bands, and the
Hall conductivity is related to the Chern number as follows,
\begin{equation}
 \sigma_{xy}=-\nu\frac{e^2}{h},\qquad
  \nu=\sum_{n\in{\rm occupied}}\nu_{n}.
\end{equation}
This is so-called TKNN formula
\cite{TKNN,Avron1983,Kohmoto1985,Niu-Thouless-Wu}.

\subsection{Fukui-Hatsugai-Suzuki's method}
Now we show that the Chern numbers in lattice systems are related to
those defined in the continuous Brillouin zone
\eqref{chern_number_continuum} in the continuum limit.  By the Taylor
expansion, $V_{\bm{k},\bm{k}+\hat{k}_{\mu}}$ defined by
Eq.~\eqref{U(1)_link} in the continuum limit becomes
\begin{equation}
 V_{\bm{k},\bm{k}+\hat{k}_{\mu}} \simeq
  1-\i\frac{2\pi}{L_{\mu}}A_{\mu}(\bm{k}),
\end{equation}
where $A_{\mu}(\bm{k})$ is the Berry connection defined in
\eqref{Berry_connection}. Similarly, we get
\begin{align}
 V_{\bm{k}+\hat{k}_{\mu},\bm{k}+\hat{k}_{\mu}+\hat{k}_{\nu}}
 &\simeq 1-\i\frac{2\pi}{L_{\nu}}A_{\nu}(\bm{k}+\hat{k}_{\mu})\\
 &\simeq 1-\i\frac{2\pi}{L_{\nu}}A_{\nu}(\bm{k})
 -\i\frac{(2\pi)^2}{L_{\mu}L_{\nu}}\partial_{\mu}A_{\nu}(\bm{k}),
 \nonumber
\end{align}
where $\mu\neq\nu$, and we have abbreviated as
$\partial_{\nu}=\partial_{k_{\nu}}$.  Next, $F_{12}(\bm{k})$ defined by
Eq.~\eqref{field_stlength} becomes
\begin{widetext}
\begin{align}
 F_{12}(\bm{k})
 &\simeq\ln\Biggl[\left(1-\i\frac{2\pi}{L_{1}}A_{1}\right)
 \left(1-\i\frac{2\pi}{L_{2}}A_{2}-\i\frac{(2\pi)^2}{L_{1}L_{2}}
 \partial_{1}A_{2}\right)
%\nonumber\\
% &\times
 \left(1+\i\frac{2\pi}{L_{1}}A_{1}
 +\i\frac{(2\pi)^2}{L_{1}L_{2}}\partial_{2}A_{1}\right)
 \left(1+\i\frac{2\pi}{L_{2}}A_{2}\right)\Biggr]\nonumber\\
 &\simeq\ln\left[1-\i\frac{(2\pi)^2}{L_{1}L_{2}}
 \bigl(\partial_{1}A_{2}-\partial_{2}A_{1}\bigr)\right]\nonumber\\[5pt]
 &\simeq-\i\frac{(2\pi)^2}{L_{1}L_{2}}
 \bigl(\partial_{1}A_{2}-\partial_{2}A_{1}\bigr),
\end{align}
so that the Chern number in lattice systems (\ref{FHS_chern}) becomes
\begin{equation}
 \nu\simeq\frac{1}{2\pi}\iint\d^2 \bm{k}
  \bigl[\partial_{1}A_{2}(\bm{k})-\partial_{2}A_{1}(\bm{k})\bigr].
\end{equation}
This coincides with the TKNN formula (\ref{chern_number_continuum}).

\subsection{Coh-Vanderbilt's method}

The relations of the Coh-Vanderbilt's (CV's) method (\ref{CV_method})
are confirmed by the following calculations,
\begin{align}
\prod_{\bm{k}}\braket{u(k_{1},k_{2}+\tfrac{\phi}{L_{2}})|
  u(k_{1}-\tfrac{2\pi}{L_1},k_{2}+\tfrac{\phi}{L_{2}})}
 &\simeq \prod_{\bm{k}}\left(1-\frac{2\pi}{L_{1}}
 \braket{u(k_{1},k_{2}+\tfrac{\phi}{L_2})
 |\partial_{1}u(k_{1},k_{2}+\tfrac{\phi}{L_2})}\right)
 \nonumber\\
 &\simeq\exp\biggl[\,\i\frac{L_2}{2\pi}
 \iint\d^2\bm{k}\,A_{1}(k_{1},k_{2}+\tfrac{\phi}{L_{2}})\biggr]
 \nonumber\\
 &\simeq\exp\biggl[\,\i\frac{L_2}{2\pi}
 \iint\d^2\bm{k}\,A_{1}(k_{1},k_{2})
 \biggr]
 \exp\biggl[\,\i\phi\underbrace{\frac{1}{2\pi}
 \iint\d^2\bm{k}\,\partial_{2} A_{1}(k_{1},k_{2})}_{-\nu}
 \biggr].
\end{align}
In this case, the Chern number is given by the Landau gauge, wheres the
FHS's method is corresponds to the symmetric gauge. Thus we get the
relation
\begin{equation}
 z(0,\phi)=z(0,0)\e^{-\i\phi\,\nu}.
\end{equation}
Similarly, we get
\begin{equation}
 \braket{\Psi_{0}(\phi,0)|U_{2}|\Psi_{0}(\phi,0)}
 \sim z(0,0)\e^{\i\phi\,\nu}.
\end{equation}
Thus we have shown that the relationship between $z(\phi)$ and the Chern
number $\nu$ are given by Eq~(\ref{chern_CV}).

However, we can modify the CV's method so that the Chern number is given
by the symmetric gauge. For this purpose, we define the polarization as
follows,
\begin{align}
 z(\phi)
 &=\braket{\Psi_{0}(\phi,\phi)|
 U_1^{\mathstrut}U_{2}^{-1}|\Psi_{0}(\phi,\phi)}\nonumber\\
 &=-\prod_{\bm{k}}
 \braket{u(k_{1}+\tfrac{\phi}{L_{1}},k_{2}+\tfrac{\phi}{L_{2}})
 |u(k_{1}+\tfrac{\phi}{L_{1}}-\tfrac{2\pi}{L_{1}},
 k_{2}+\tfrac{\phi}{L_{2}}+\tfrac{2\pi}{L_{2}})}.
 \label{flux_chern_number2}
\end{align}
Then \eqref{flux_chern_number2} becomes
\begin{equation}
 z(\phi_{1},\phi_{2})
  =-\prod_{\bm{k}}
  \braket{u(k_{1}+\tfrac{\phi_1}{L_{1}},k_{2}+\tfrac{\phi_2}{L_{2}})|
  u(k_{1}+\tfrac{\phi_{1}-2\pi}{L_{1}},k_{2}+\tfrac{\phi_{2}+2\pi}{L_{2}})}.
\end{equation}
In the thermodynamic limit $L_{\mu}\gg1$, the Taylor expansion gives
\begin{align}
 z(0,0)&=-
 \prod_{\bm{k}}\braket{u(k_{1},k_{2})|
 u(k_{1}-\tfrac{2\pi}{L_{1}},k_{2}+\tfrac{2\pi}{L_{2}})}\nonumber\\
 &\simeq-\exp\left[\i
 \iint\d^2\bm{k}
 \left\{
 \frac{L_{2}}{2\pi}
 A_{1}(k_{1},k_{2})
 -
 \frac{L_{1}}{2\pi}
 A_{2}(k_{1},k_{2})
 \right\}
 \right],
\end{align}
where $A_{\mu}(k_{1},k_{2})$ is a component of the Berry connection
defined in Eq.~\eqref{Berry_connection}.  Therefore
Eq.~\eqref{flux_chern_number2} becomes
\begin{align}
 z(\phi_{1},\phi_{2})
 &=-\exp\left[\i
 \iint\d^2\bm{k}
 \left\{
 \frac{L_{2}}{2\pi}
 A_{1}(k_{1}+\tfrac{\phi_{1}}{L_{1}},k_{2}+\tfrac{\phi_{2}}{L_{2}})
 -
 \frac{L_{1}}{2\pi}
 A_{2}(k_{1}+\tfrac{\phi_{1}}{L_{1}},k_{2}+\tfrac{\phi_{2}}{L_{2}})
 \right\}
 \right]\nonumber
 \\[5pt]
 &=z(0,0)
 \exp\left[\i
 \iint\d^2\bm{k}
 \left\{
 \frac{L_{2}}{L_{1}}\frac{\phi_{1}}{2\pi}\partial_1A_{1}(\bm{k})
 +
% \frac{L_{2}}{2\pi}\frac{\phi_{2}}{L_{2}}\partial_2A_{1}(\bm{k})
 \frac{\phi_{2}}{2\pi}\partial_2A_{1}(\bm{k})
 -
% \frac{L_{1}}{2\pi}\frac{\phi_{1}}{L_{1}}\partial_1A_{2}(\bm{k})
 \frac{\phi_{1}}{2\pi}\partial_1A_{2}(\bm{k})
 -
 \frac{L_{1}}{L_{2}}\frac{\phi_{2}}{2\pi}\partial_2A_{2}(\bm{k})
 \right\}
 \right]
 \nonumber
 \\
 &\simeq z(0,0)\exp\left[-\i\frac{\phi}{2\pi}
 \iint\d^2\bm{k}
 (\partial_1A_{2}(\bm{k})-\partial_2A_{1}(\bm{k}))
 \right],\qquad(\phi=\phi_1=\phi_2,\, L_1=L_2)
 \nonumber
 \\[5pt]
 &=z(0,0)\exp(-\i\phi\,\nu).
\end{align}
Thus we have obtained the Chern number given by the symmetric gauge.

\end{widetext}

%%%%%%%%%%%%%%%%%%%%%%%%%%%%%%%%%%%%%%%%%%%%%%%%%%%%%%%%%%%%%%%%%%%%%%
% References
%%%%%%%%%%%%%%%%%%%%%%%%%%%%%%%%%%%%%%%%%%%%%%%%%%%%%%%%%%%%%%%%%%%%%%

\end{document}